\documentclass[11pt]{article}
\usepackage{amsmath}
\usepackage{amsthm}
\usepackage{amssymb}
\usepackage{amsfonts}
\usepackage{mathtools}
\usepackage{bbm}
\usepackage{bm}
\usepackage{booktabs}
\usepackage{graphicx}
\usepackage{subfigure}
\usepackage{multirow}
\usepackage{longtable}
\usepackage{adjustbox}
\usepackage{longtable}
\usepackage{algorithm}
\usepackage{algorithmic}
\usepackage{microtype}
\usepackage{xspace}
\usepackage{verbatim}
\usepackage{color}
\usepackage{comment}

\usepackage{natbib}
\usepackage{hyperref}
\usepackage[capitalize,noabbrev]{cleveref}
\usepackage{url}

\usepackage{authblk}

\makeatletter
\g@addto@macro{\UrlBreaks}{\UrlOrds}
\makeatother
\usepackage{threeparttable}

\newtheorem{Theorem}{Theorem}
\newtheorem{Lemma}{Lemma}
\newtheorem{assumption}{Assumption}
\usepackage[textsize=tiny]{todonotes}
\usepackage{pdflscape}

\def\bSig\mathbf{\Sigma}

\renewcommand{\mid}{\mbox{\,$|$\,}}

\title{Online multi-layer FDR control}

\author[1]{Runqiu Wang}
\author[1]{Ran Dai\thanks{ran.dai@unmc.edu}}

\affil[1]{Department of Biostatistics, University of Nebraska Medical Center, Omaha, Nebraska, U.S.A.}

\begin{document}

\maketitle



\begin{abstract}
When hypotheses are tested in a stream and real-time decision-making is needed, online sequential hypothesis testing procedures are needed. Furthermore, these hypotheses are commonly partitioned into groups {\color{black}by their nature}. For example, the RNA nanocapsules can be partitioned based on therapeutic nucleic acids (siRNAs) being used, as well as the delivery nanocapsules. When selecting effective RNA nanocapsules, simultaneous false discovery rate control at multiple partition levels is needed. In this paper, we develop hypothesis testing procedures which controls false discovery rate (FDR) simultaneously for multiple partitions of hypotheses in an online fashion. We provide rigorous proofs on their FDR or modified FDR {\color{black}(mFDR)} control properties and use extensive simulations to demonstrate their performance.

%
%

\end{abstract}

\section{Introduction}
In early stage drug development research, lead compounds are often selected in an online fashion to enter the next-stage experiment (i.e. from discovery to in vitro study, or from in vitro to in vivo study)\citep{deore2019stages}. Because the data enters in a stream or in batches, and real-time decision-makings are desirable, online sequential hypothesis testing procedures need to be adopted. Furthermore, The hypotheses in drug discovery are commonly partitioned into groups by their nature, and multiple ways of partition are simultaneously of interest. For example, in developing innovative RNA nanocapsules to treat cancer metastasis, therapeutic nucleic acids (siRNAs) as well as delivery nanocapsules are designed \citep{lee2013recent,ebrahimi2023nano}. At each step of the screening, a decision needs to be made for whether to follow up on a particular combination of a siRNA and nanocapsule. In this case, these combinations are naturally partitioned according to the siRNA and nanocapsule formulations. Ideally, to selecting the candidate lead compounds, we are interested in controlling the false positives in terms of both siRNA and nanocapsule formulations. Potentially, we are interested in other partitions, for example, based on the experiments with different cancer types \citep{zhang2008hit}. 

{\color{black}Beyond drug development, the need for multi-layer false discovery rate (FDR) control also arises in many modern applications. In real-time A/B testing on online platforms, experiments are often run concurrently across user segments, device types, or product categories, and results arrive continuously \citep{taddy2015heterogeneous, xie2018false}. Similar challenges also exist in genomics, where hypotheses (e.g., gene expressions or mutations) may be partitioned by biological pathways or genomic regions, and tested sequentially in large-scale screening studies \citep{goeman2007analyzing}. These situations require statistical methods that can make decisions in an online fashion, while still simultaneously control the FDR across multiple layers of structures. Motivated by these examples, we aim to develop a hypothesis-testing procedure that controls the FDR simultaneously for multiple partitions of hypotheses in an online fashion. 
}

Mathematically, at time $t \in \mathbb{N}$, we have observed a stream of null hypotheses $H_{i(t)j(t)}$'s, where $i(t) \in \mathbb{N}$ denotes the index for the siRNA being tested at time $t$, and $j(t) \in \mathbb{N}$ denotes the index for the nanocapsule being tested at time $t$. $H_{i(t)j(t)}$ indicates a hypothesis with respect to the siRNA $i(t)$ and nanocapsule $j(t)$.

We define two partition-level null hypotheses $H^1_{i(t)}$ and $H^2_{j(t)}$ such that
\[H^1_{i(t)} ~\text{is true}~ \iff  H_{i(t)j} ~\text{is true for}~ j \in \cup_{u = 1}^t\{j(u)\},\]
\[H^2_{j(t)} ~\text{is true}~ \iff  H_{i{j(t)}} ~\text{is true for}~ i \in \cup_{u = 1}^t\{i(u)\}  ,\]
where $H^1_{i(t)}$ is the partition-level hypothesis with respect to the siRNAs, and $H^2_{i(t)}$ is the partition-level hypothesis with respect to the nanocapsules,
 {\color{black} $H_{i(t)j}$ is the hypothesis where the siRNA index $i(t)$ is fixed at time t, and we vary over nanocapsule indices $j$, $H_{ij(t)}$ is a hypothesis where the nanocapsule index $j(t)$ is fixed at time t, and vary over siRNA indices $i$.}
At time $t$, we make a decision on whether to reject $H_{i(t)j(t)}$ based on its corresponding test statistic and p-value, as well as the history information on the previous hypotheses (statistic, p-value and decisions), i.e. we let $\delta_t = 1$ if we reject $H_{i(t)j(t)}$, and $\delta_t = 0$ if we accept $H_{i(t)j(t)}$. Our goal is to control the group-level FDR with both partitions according to siRNA or according to nanocapsules, which we denote as $\text{FDR}^1(t)$ and $\text{FDR}^2(t)$. Let 
\[\widehat{S}^1(t) = \{i(u): \delta_u = 1\}, \widehat{S}^2(t) = \{j(u): \delta_u = 1\},\]
\[\text{FDR}^1(t) = \mathbb{E}\left[\frac{|\widehat{S}^1(t) \cap H^1_{i(t)}|}{|\widehat{S}^1(t)| \vee 1} \right],  \text{FDR}^2(t) = \mathbb{E}\left[\frac{|\widehat{S}^2(t) \cap H^2_{j(t)}|}{|\widehat{S}^2(t)| \vee 1} \right],\]
{\color{black}where $\widehat{S}^1(t)$ and $\widehat{S}^2(t)$ denote the sets of rejected hypotheses at time $t$ grouped by siRNA and nanocapsule partitions, respectively; $H_{i(t)}^1$ and $H_{j(t)}^2$ denote the corresponding sets of true null hypotheses for the siRNA and nanocapsule partitions, respectively. $\vee$ is the maximum operator: $|\widehat{S}^1(t)| \vee 1=\max(|\widehat{S}^1(t)|,1)$.}

{\color{black}To control FDR for individual online hypotheses, Foster and Stine (2008)\cite{foster2008alpha} introduced the $\alpha$-investing framework, a foundational method for online FDR control. This approach treats significance thresholds as a budget, namely $\alpha$-wealth, which is spent to test hypotheses and ``spent" upon discoveries (i.e., rejected null hypotheses). This method controls the modified FDR (mFDR) and is sensitive to test ordering. {\color{black}The modified false discovery rate (mFDR) is an alternative measure defined by the ratio between the expected number of false discoveries and the expected number of discoveries.} To address these limitations, Javanmard and Montanari (2018)\cite{javanmard2018online} proposed two alternative procedures: LOND (Levels based On Number of Discoveries) and LORD (Levels based On Recent Discoveries) to provide finite-sample FDR control under dependency structures. In LOND, the significance level at each step is scaled according to the number of previous rejections. LORD, on the other hand, updates significance thresholds after each discovery, allowing for greater adaptability. Both methods depends on independent p-values. More recently, these methods have been further extended for improved power and relaxed dependence assumptions for the p-values \citep{ramdas2018saffron, tian2021online, zehetmayer2022online}. 

For the FDR control in grouped or structured hypotheses, such as genes grouped by biological pathways or drug compounds tested across cancer types, Ramdas and Barber (2019)\citep{ramdas2019unified} developed the p-filter, a method for multi-partitioned data settings where hypothesis testing occurs across distributed or federated environments. p-filter maintains a dynamic filtration level that propagates across data partitions, enabling asynchronous hypothesis testing while preserving FDR guarantees. Ramdas et al. (2019) \citep{ramdas2019sequential} developed the DAGGER method, which controls the FDR in hypothesis testing structured as directed acyclic graphs, where nodes represent hypotheses with parent-child dependencies. DAGGER processes hypotheses top-down, enforcing that a node is rejected only if its parents are rejected, and guarantees FDR control under independence, positive, or arbitrary dependence of p-values.}


{\color{black} In many practical applications drug discovery, as we described in the introduction, there is a need to control the FDR at multiple partition levels simultaneously in an online fashion; there is no existing method for this kind of settings. We aim to address this gap by developing a FDR control method to control group-level FDR defined based on multiple structures or partitions (multiple layers) in an online fashion. In our setting, not only the hypotheses enter sequentially, but the group membership of the entering hypothesis in each layer is also not predefined. We develop a novel framework that integrates the online FDR control in multi-layer structured testing capabilities to provide theoretically guaranteed FDR control in multi-layer online testing environments. The main contributions of this paper are summarized below:
\begin{itemize}
\item We present multi-layer $\alpha$-investing, LOND and LORD algorithms for controlling FDR in online multiple-layer problems.
\item We provide the theoretical guarantees for multi-layer $\alpha$-investing, LOND and LORD methods.
\item We demonstrate the numerical FDR control performance and the power of our methods with extensive simulation settings. 
\end{itemize}

The rest of the paper is structured as follow. In Section \ref{model}, we provide a mathematical formulation of the online multi-layer hypothesis testing problem. In Section \ref{algorithms}, we illustrate multi-layer $\alpha$-investing, LOND and LORD algorithms. In Section \ref{result}, we show our main theoretical results, that multi-layer $\alpha-$ investing method simultaneously controls mFDR at different layers; and the multi-layer LOND and LORD simultaneously control FDR and mFDR at different layers. In Section \ref{simulation}, we demonstrate the performance of our methods with extensive simulation studies.}


\section{Model Framework}\label{model}

Now we formulate the problem for more general scenarios. From now on, we use the following notations. Suppose we have a series of (unlimited) null hypotheses $H_i^0$s, $i \in \mathbb{N}$ and $M$ partitions of the space $\mathbb{N}$, denoted as $A_1^m, \cdots, A_{G_m}^m \subseteq \mathbb{N}$ for $m = 1, \cdots, M$. We call these $M$ different partitions ``layers" of hypotheses. We can define group-level null hypotheses in layer $m$ at time $t$ as \[H_{A_g^m}^0(t) = \cap_{j\in A_g^m \cap [t]}H_j^0, ~\text{where}~g = 1, \cdots, G_m,\]{\color{black}where [t] = \{1,2,\dots,t\}}. We also define a group index for the $i$-th hypothesis in the $m$-th layer as $g_i^m$, i.e. $i \in A_{g_i^m}^m$.

In the online setting, $H_i^0$s come in sequentially and the corresponding test statistics $z_i$'s and associated p-values $p_i$'s are also observed sequentially. For each $H_i^0$ we need to make the decision $\delta_i$ on whether to reject it based on $z_i$, $p_i$ and the available history information, i.e. $\delta_i = 1$ if we reject $H_i^0$, $\delta_i = 0$ if we accept $H_i^0$. At time $t$ (t is discretized), we have three different types of history information: 
\begin{enumerate}
    \item All information $\mathcal{F}_t^a = \{z_i, p_i, \delta_i, i = 1, \cdots, t\}$.
    \item All decisions $\mathcal{F}_t^d = \{\delta_i, i = 1, \cdots, t\}$.
    \item Summary of decisions, for example $D_t = \sum_{i=1}^t \delta_i$.
\end{enumerate}

Now, we define decisions at group level using decisions at individual level. At individual level, we define the rejection set at time $t$ as $\widehat{S}(t) = \{i : \delta_i = 1\} \cap [t]$. Since the decision whether to reject a group will change over time, we use $\delta_g^m(t)$ to denote the decision on group $g$ ($g \in \{1, \cdots, G_m\}$) in layer $m$ at time $t$ where \[\delta_g^m(t) = 1 ~\text{if}~ \widehat{S}(t) \cap A_g^m \neq \emptyset ~\text{and}~ \delta_g^m(t) = 0 ~\text{otherwise}.\] So for the $m$-th partition, the group level selected sets at time $t$ are \[\widehat{S}^m(t) = \{g \in \{1, \cdots, G_m\} : \delta_g^m(t) = 1\}.\] 
Similarly, denote the truth about $H_i$ at individual level by $\theta_i$; $\theta_i = 0$ when $H_i^0$ should be accepted while $\theta_i = 1$ when $H_i^0$ should be rejected. We can obtain the true set $S(t) = \{i : \theta_i = 1\} \cap [t]$ and \[\text{$\theta_g^m(t) = 1$ if $S(t) \cap A_g^m \neq \emptyset$ and $\theta_g^m(t) = 0$ otherwise.}\] 

For the $m$-th partition, the group-level true null set at time $t$ is $S^m(t) = \{g \in \{1, \cdots, G_m\} : \theta_g^m(t) = 1\}$.

Now we can define the time-dependent false discovery rate (FDR) and modified false discovery rate (mFDR) for each layer separately. For the $m$-th layer, we have \[\text{FDR}^m(t) = \mathbb{E}[\frac{V^m(t)}{R^m(t) \vee 1}] ~\text{and}~ \text{mFDR}_{\eta}^m(t) = \frac{\mathbb{E}V^m(t)}{\mathbb{E}R^m(t)+\eta},\] where $V^m(t) = |\widehat{S}^m(t) \cap (S^m(t))^c|$ and $R^m = |\widehat{S}^m(t)|$.

\section{Algorithms}\label{algorithms}

\subsection{Multi-layer $\alpha$-investing}

In this section, we illustrate the generalized $\alpha$-investing algorithm for the multi-layer setting.

A multi-layer $\alpha$-investing procedure is an adaptive sequential hypothesis testing procedure. Suppose the overall target FDR is $\alpha$ for every layer. For each time point $t$ ($t = 1, \cdots, N$ where we allow $N \to \infty$) and partition layer $m$ ($m = 1, \cdots, M$), we define the significance level $\alpha_t^m$, the spending function $\phi_t^m$, and rewarding function $\psi_t^m$ that satisfy the assumptions in Theorem 1 and Lemma 1 in Section 3 (a simple choice is alpha-investing \cite{foster2008alpha} procedure with $\alpha_t^m = \alpha$, $\phi_t^m = \frac{\alpha}{1-\alpha}$, $\psi_t^m = \phi_t^m + \alpha$). For each hypothesis $H_i^0$, it will be rejected if its p-values $p_i^m$'s pass the corresponding thresholds $\alpha_i^m$'s for all those $m$'s that are still pending for a decision (that is, $\forall m$'s s.t. $\delta_{g_i^m}^m(i - 1) = 0$). The wealth functions $W^m(t)$ is initialized at $\alpha\eta$ and then updated under the following rule: at time $t = i$, if the hypothesis $H_i^0$ is rejected, then we gain reward $\psi_i^m$ for all the layers tested at this round. In the mean time, no matter whether $H_i^0$ is rejected, we lose spending $\phi_i^m$ for those layers that are tested in this round. In other words, we can write $W^m(i)$ as:
\[
W^m(i) = 
\begin{cases}
W^m(i - 1) + \psi_i^m - \phi_i^m, & \delta_{g_i^m}^m(i - 1) = 0, \delta_{g_i^m}^m(i) = 1 \\
W^m(i - 1) - \phi_i^m, & \delta_{g_i^m}^m(i - 1) = 0, \delta_{g_i^m}^m(i) = 0
\end{cases}
\]

This procedure simultaneously controls mFDR at all layers. We state the multi-layer $\alpha-$ investing algorithm in Algorithm 1.

\begin{algorithm}
\caption{Multi-layer $\alpha-$ investing}
\begin{algorithmic}[1]
\STATE Initialize $\delta_g^m = 0$ and $W^m(0) = \alpha\eta$, for $m = 1, \cdots, M$ and all $g \in \{1, \cdots, G_m\}$.
\WHILE{$\min_m W^m(i) > 0$ and $i \leq N$}
\STATE $\delta_i = 1$
\FOR{$m = 1, \cdots, M$}
\IF{$\delta_{g_i^m}^m = 0$}
\IF{$p_i^m \geq \alpha_i^m$}
\STATE $\delta_i = 0$, break
\ENDIF
\ENDIF
\ENDFOR
\IF{$\delta_i = 1$}
\FOR{$m = 1, \cdots, M$}
\IF{$\delta_{g_i^m}^m = 0$}
\STATE $\delta_{g_i^m}^m = 1$ and $W^m(i) = W^m(i - 1) + \psi_i^m - \phi_i^m$
\ENDIF
\ENDFOR
\ELSE
\FOR{$m = 1, \cdots, M$}
\IF{$\delta_{g_i^m}^m = 0$}
\STATE $W^m(i) = W^m(i - 1) - \phi_i^m$
\ENDIF
\ENDFOR
\ENDIF
\STATE $i=i+1$
\ENDWHILE
\end{algorithmic}
\end{algorithm}

\textbf{Remark:} The statistics $p_i^m$'s are also allowed to be just individual $p_i$'s for all layers or a combined score, which is a function of all $p_j$'s s.t. $j \leq i$ and $g_j^m = g_i^m$.

\subsection{Multi-layer LOND and LORD}

In this section, we generalize the LOND and LORD algorithms for the multi-layer setting.

To extend the LOND algorithm for the multi-layer setting, we set the significance level for the hypothesis at time $i$ in layer $m$ to be $\beta_i^m(R^m(i - 1) + 1)$, where \[\left\{\beta_i^m : \sum_{i=1}^{\infty} \beta_i^m = \alpha \right\} ~\text{for}~m = 1, \cdots, M\] are $M$ predefined infinite sequences that sum to 1, and $R^m(i - 1) = |\widehat{S}^m(i - 1)|$. For each hypothesis $H_i^0$, it will be rejected if its p-value $p_i^m$ passes the threshold $\beta_i^m(R^m(i - 1) + 1)$ for all those $m$'s that are still pending for a decision ($\forall m$'s s.t. $\delta_{g_i^m}^m(i - 1) = 0$). We state the mulit-layer LOND algorithm in Algorithm 2 and we will show that the multi-layer LOND algorithm can simultaneously control multi-layer FDR and mFDR at level $\alpha$ in the next section.

\begin{algorithm}
\caption{Multi-layer LOND}
\begin{algorithmic}[1]
\STATE Select sequences $\sum_{j=1}^{\infty} \beta_j^m = \alpha$ and initial $R^m = 0$, and $\delta_g^m = 0$, for $m = 1, \cdots, M$ and all $g \in \{1, \cdots, G_m\}$.
\WHILE{$i \leq N$}
\STATE $\delta_i = 1$
\FOR{$m = 1, \cdots, M$}
\IF{$\delta_{g_i^m}^m = 0$}
\IF{$p_i^m \geq \beta_i^m(R^m + 1)$}
\STATE $\delta_i = 0$, break
\ENDIF
\ENDIF
\ENDFOR
\IF{$\delta_i = 1$}
\FOR{$m = 1, \cdots, M$}
\IF{$\delta_{g_i^m}^m = 0$}
\STATE $R^m = R^m + 1$, $\delta_{g_i^m}^m = 1$
\ENDIF
\ENDFOR
\ENDIF
\STATE $i=i+1$
\ENDWHILE
\end{algorithmic}
\end{algorithm}

\textbf{Remark:} Since in layer $m$, we do not need to test a hypothesis at every time point $t = 1, \cdots, i$, in fact, we only did $\kappa_i^m$ tests, where $\kappa_i^m = i-\sum_{j=1}^{i}\delta_{g_j^m}^m(i)+\sum_{j=1}^{G_m}\delta_j^m(i)$, we can improve the method by setting the significance level for the hypothesis at time $i$ in layer $m$ to be $\beta_{\kappa_i^m}^m(R^m + 1)$. By doing this, the power is increased and the multi-layer FDRs are still controlled. We will show this improvement by simulation in Section 4, but omit the theoretical proof of it.

To extend the LORD algorithm for the multi-layer setting, we set $M$ predefined sequences $\{\beta_j^m : \sum_{j=1}^{\infty} \beta_j^m = \alpha\}$. For the hypothesis at time $i$ in layer $m$ we use significance level $\alpha_i^m = \beta_{\zeta_i^m}^m$ where $\zeta_i^m = \kappa_i^m -\kappa_{\tau_i^m}^m + 1$, $\tau_i^m = \max\{j < i : \delta_{g_j^m}^m(j - 1) = 0, \delta_{g_j^m}^m(j) = 1\}$ and $\kappa_i^m = i-\sum_{j=1}^{i}\delta_{g_j^m}^m(i)+\sum_{j=1}^{G_m}\delta_j^m(i)$. Define $\zeta_i^m = \kappa_i^m -\kappa_{\tau_i^m}^m + 1$. We can show that the modified LORD algorithm simultaneously controls multi-layer FDR and mFDR at level $\alpha$.

\begin{algorithm}
\caption{Multi-layer LORD}
\begin{algorithmic}[1]
\STATE Select sequences $\sum_{g=1}^{\infty} \beta_g^m = \alpha$ and initial $\zeta^m = 1$
\WHILE{$i \leq N$}
\STATE $\delta_i = 1$
\FOR{$m = 1, \cdots, M$}
\IF{$\delta_{g_i^m}^m = 0$}
\IF{$p_i^m \geq \beta_{\zeta^m}^m$}
\STATE $\delta_i = 0$, break
\ENDIF
\ENDIF
\ENDFOR
\IF{$\delta_i = 1$}
\FOR{$m = 1, \cdots, M$}
\IF{$\delta_{g_i^m}^m = 0$}
\STATE $\zeta^m = 1$ and $\delta_{g_i^m}^m = 1$
\ENDIF
\ENDFOR
\ELSE
\FOR{$m = 1, \cdots, M$}
\IF{$\delta_{g_i^m}^m = 0$}
\STATE $\zeta^m = \zeta^m + 1$
\ENDIF
\ENDFOR
\ENDIF
\STATE $i=i+1$
\ENDWHILE
\end{algorithmic}
\end{algorithm}

\section{Theoretical Results}\label{result}

We will show that multi-layer $\alpha-$ investing method simultaneously controls mFDR at different layers; and the multi-layer LOND and LORD simultaneously control both FDR and mFDR at different layers. 
{\color{black}
First we introduce one preliminary assumption, this assumption is weaker than the independent assumption across the tests. 
}
\begin{assumption}
$\forall m \in \{1, \cdots, M\}$, and all layers that need to be tested at time $j$,
\begin{align*}
\forall \theta_j = 0 &: P_{\theta}(\delta_{g_j^m}^m(j) = 1|\mathcal{F}_{j-1}) \leq \alpha_j^m, \\
\forall \theta_j = 1 &: P_{\theta}(\delta_{g_j^m}^m(j) = 1|\mathcal{F}_{j-1}) \leq \rho_j^m.
\end{align*}
\end{assumption}

{\color{black}
\textbf{Remark:} This assumption is a simple extension of Assumption 2.2 of the generalized $\alpha$-investing \cite{javanmard2015online}. It requires  that the level and power traits of a test hold even given the history of rejections.
This is somewhat weaker than requiring independence. In particular, $\rho_j^m \leq 1, \forall m \in \{1,\cdots,M\}$ is the maximal probability of rejection for any parameter value in the alternative hypothesis, which quantifies the ``best power".

Next we state the theoretical $mFDR$ control guarantee for the milti-layer generalized $\alpha$-investing procedure.
}

\begin{Theorem}\label{method:theorem1}
Under Assumption 1, if $0 \leq \psi_j^m \leq \min(\frac{\phi_j^m}{\rho_j^m} + \alpha, \frac{\phi_j^m}{\alpha_j^m} + \alpha + 1)$, then a multi-layer generalized $\alpha$-investing procedure simultaneously control $mFDR_{\eta}^m$ for all layers $m = 1, \cdots, M$.
\end{Theorem}

{\color{black} The proof of Theorem \ref{method:theorem1} is postponed to Appendix A. One key step is to establish a submartingale as stated in the following lemma.}

\begin{Lemma}\label{method:lemma1}
Under Assumptions 1, and the multi-layer $\alpha$-investing procedure, the stochastic processes $A^m(j) = \alpha R^m(j) - V^m(j) + \alpha \eta - W^m(j)$, $m = 1, \cdots, M$ are submartingales with respect to $\mathcal{F}_j$.
\end{Lemma}

{\color{black}
Next, we state the $FDR$ and $mFDR$ control guarantees for the multi-layer LOND and LORD methods.}

\begin{Theorem}\label{method:theorem2}
Suppose that conditional on \( R^m(i - 1) = |\widehat{S}^m(i - 1)| \), \( m = 1, \dots, M \), we have
\[
\forall \theta_i \in \Theta, \forall m = 1, \dots, M,
\]
\[
\mathbb{P}_{\theta_i = 0} \left( \delta^m_{g_i^m}(i) = 1 \middle| R^m(i - 1), \delta^m_{g_i^m}(i - 1) = 0 \right) \leq \mathbb{E} \left( \alpha_i^m \middle| R^m(i - 1), \delta^m_{g_i^m}(i - 1) = 0 \right),
\]
where \( \alpha_i^m = \beta_i^m (R^m(i - 1) + 1) \) and \( \delta_i^m \) is given by the LOND algorithm. Then this rule simultaneously controls multi-layer FDRs and mFDRs at level less than or equal to \( \alpha \), i.e.,
\[
\forall n \geq 1, \forall m = 1, \dots, M, \quad \mathrm{FDR}^m(n) \leq \alpha \quad \text{and} \quad \mathrm{mFDR}^m(n) \leq \alpha.
\]
\end{Theorem}

\begin{Theorem}\label{method:theorem3}
Suppose that conditional on the history of decisions in layer $m$, 
$
\mathcal{F}_t^m = \sigma\left\{ \delta_{g_i^m}^m(i), i = 1, \ldots, t \right\}
$
we have $\forall \theta \in \Theta, \forall m = 1, \ldots, M$ that need to be tested at time $i$,
\[
\mathbb{P}_{\theta_i = 0}\left( \delta_{g_i^m}^m(i) = 1 \mid \mathcal{F}_t^m \right) \leq \mathbb{E}\left( \alpha_i^m \mid \mathcal{F}_t^m \right).
\]

Then, the LORD rule simultaneously controls multi-layer mFDR to be less than or equal to $\alpha$, i.e.,
$
\mathrm{mFDR}_1^m(n) \leq \alpha \quad \text{for all } n \geq 1.
$
Further, it simultaneously controls multi-layer FDR at every discovery. More specifically, letting $d_k^m$ be the time of the $k$-th discovery in layer $m$, we have the following for all $k \geq 1$,
\[
\sup_{\theta \in \Theta} \mathbb{E}_\theta \left( \mathrm{FDP}^m(d_k^m) \cdot \mathbb{I}(d_k^m < \infty) \right) \leq \alpha, \quad \forall m = 1, \ldots, M.
\]
\end{Theorem}

{\color{black}All the proofs of Lemma \ref{method:lemma1}, Theorem \ref{method:theorem1}, \ref{method:theorem2} and \ref{method:theorem3} are in Web Appendix A.}
\section{Numerical Studies}\label{simulation}

{\color{black}We conducted extensive simulations to evaluate the performance of our proposed multi-layer methods in various scenarios. We simulated data from a two-layer model with different settings to evaluate the proposed methods. For each setting, we compare the FDR and mFDR control and power at both group and individual levels. For all simulations, we set the target FDR level at $\alpha = 0.1$ and run 100 repeats.
}


\subsection{Simulation of the group structures}
{\color{black}
We consider three types of group structures: the first two are balanced, while the third is unbalanced. For the balanced strucutre, we denoted $G$ as the number of groups, and the number of hypotheses per group as $n$. Therefore, the total number of features is $N = nG$. The three types of group structures are described below:
\begin{enumerate}
\item \textbf{Balanced Block Structure}: The features comes in group by group, i.e. all features from the first groups come in, then follows all features from the second group, and so on.
\item \textbf{Balanced Interleaved Structure}: The first feature is from group 1, the second feature is from group 2,$\cdots$, the $n$th feature is from group $n$, the $(n + 1)$th feature is from group 1, and so on.
\item \textbf{Unbalanced Structure}: The features are randomly assigned to be in either the current group or a new group by some random mechanism. To be more specific, we fix the total number of features as $N$, and denote the membership of the $t$th feature as $G(t)$, let \[\mathbf{P}(G(t) = g|G(t - 1) = g) = 1 - p_1, ~~\mathbf{P}(G(t) = g'|G(t - 1) = g) = \frac{p_1}{G-1},\] where $g' \neq g \in \{1, \cdots, G\}$.
\end{enumerate}
}

\subsection{Simulation of true signals}

We consider three signal structures:
\begin{enumerate}
    \item {\color{black}\textbf{Fixed pattern:}} The first $s\%$ groups are true groups. Within the true groups, the first $k\%$ features are the true features.
    \item {\color{black}\textbf{Random pattern:}} The true groups and true signals are randomly sampled. We first randomly sample $s\%$ of the total groups to be true groups. Then within each true group, we randomly sample $k\%$ features as the true features.
    \item {\color{black}\textbf{Markov pattern:}} The signals are from a hidden Markov model. We consider a Markov model with two hidden states: the stationary state, and the eruption state. In the stationary state, $\theta_i \sim Binomial(N, 0.5)$; in the erruption state, $\mathbf{P}(\theta_i = 0|\theta_{i-1} = 0) = \mathbf{P}(\theta_i = 1|\theta_{i-1} = 1) = 0.9$. At each step, the probability that it remains in the same state as the previous step is 0.9, and it has probability of 0.1 to change to the other state.
\end{enumerate}

\subsection{Simulation of signal strength}

We consider three types of signal strength settings. 
\begin{enumerate}
    \item {\color{black}\textbf{Constant strength:}} The signal strengths are equal; the p-values for the true signals are from a two-sided test with z-statistics from $\mathcal{N}(1.5\beta, 1)$. 
    \item {\color{black}\textbf{Increasing strength:}} The signal strengths are increasing; the p-values are from two-sided test with z-statistics from $\mathcal{N}(\beta_t, 1)$, where $\beta_t = \beta(1 + t/t_{TOL})$. Here $t$ is the number of true signals until now, and $t_{TOL}$ is the total number of true signals in the simulation.
    \item {\color{black}\textbf{Decreasing strength:}} The signal strengths are decreasing; the p-values are from two-sided test with z-statistics from $N(\beta_t, 1)$, where $\beta_t = \beta(2 - t/t_{TOL})$, with $t$ and $t_{TOL}$ defined the same as in (2).
\end{enumerate}


{\color{black}All the simulation designs satisfy Assumption 1 by construction. Specifically, under the null hypothesis ($\theta_i = 0$), p-values are generated independently from the Uniform$(0,1)$ distribution. This guarantees that the probability of rejection at each step is less than or equal to the target significance level $\alpha_i^m$ (i.e., $\mathbb{P}(p_j \leq \alpha_j^m) = \alpha_j^m$), satisfying the first condition of Assumption 1. Under the alternative hypothesis ($\theta_i = 1$), p-values are generated from two-sided tests using normally distributed test statistics with non-zero means. These distributions are stochastically smaller than the null distribution, which leads to a higher probability of rejection. However, this probability is still finite and bounded above by one, ensuring that the second condition of Assumption 1 is satisfied with some constant $\rho_i^m \leq 1$. Moreover, our simulation generates data sequentially and independently, making the conditional probability bounds in Assumption 1 valid under the filtration $\mathcal{F}_{i-1}$. Therefore, both parts of Assumption 1 are met in all simulation settings, ensuring that the theoretical guarantees in Theorems 1–3 are applicable.}

{\color{black} We compare 7 methods for each setting: original $\alpha-$investing (GAI), original LORD (LORD), original LOND (LOND), multi-layer $\alpha-$investing (ml-GAI), multi-layer LORD (ml-LORD), multi-layer LOND (ml-LOND),  and the modified multi-layer LOND (ml-LOND\_m).}
\subsection{Simulation results}

\begin{figure}[H]
    \centering
     \includegraphics[scale=1]{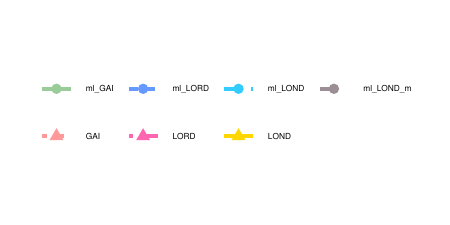}\\
     \includegraphics[scale=0.4]{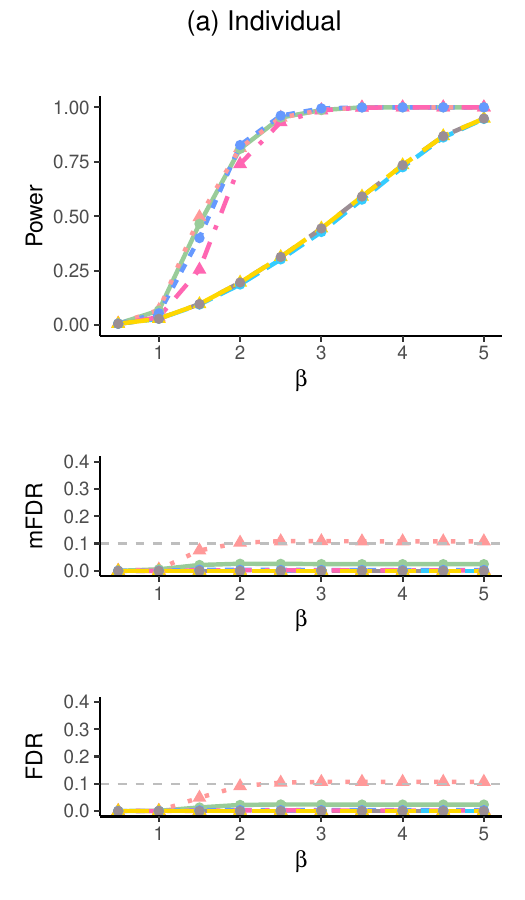}
     \includegraphics[scale=0.4]{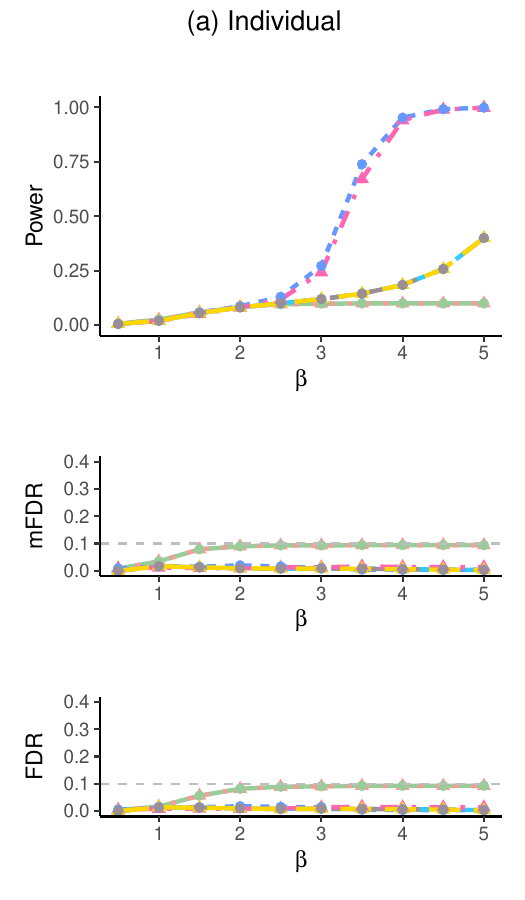}
     \includegraphics[scale=0.4]{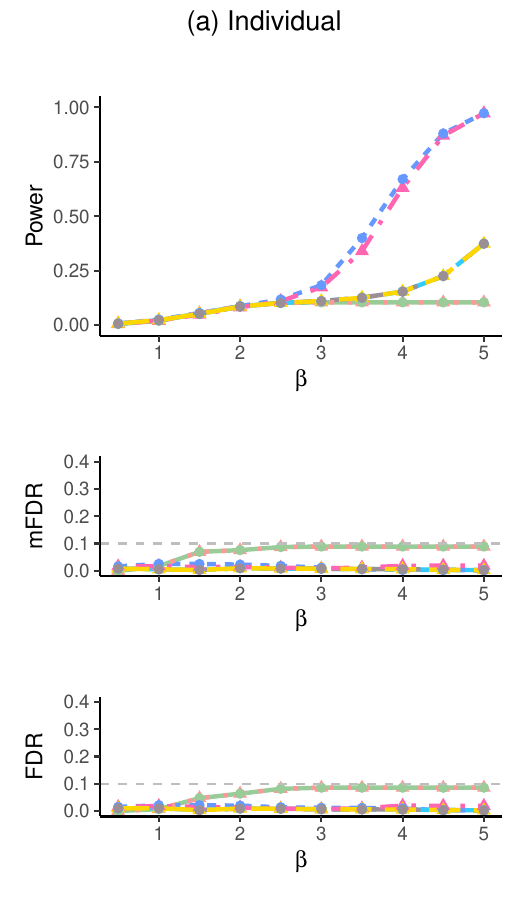}\\
     \includegraphics[scale=0.4]{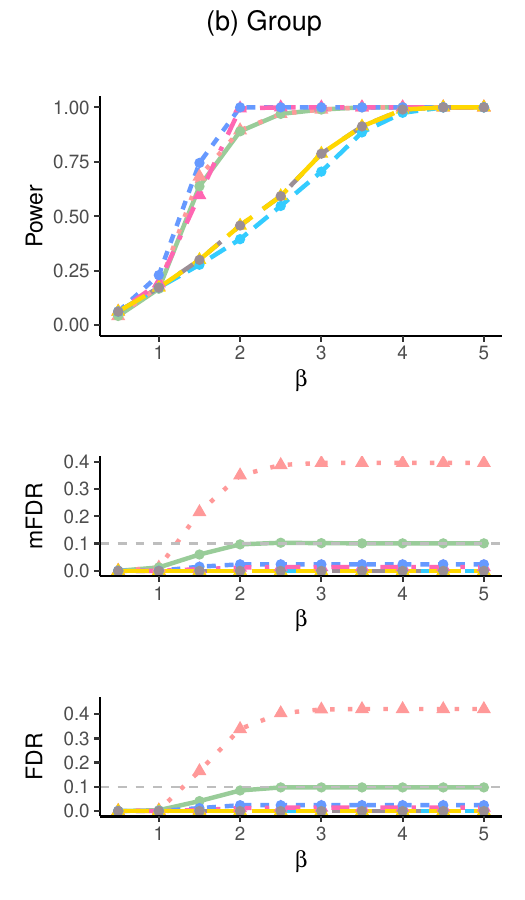}
     \includegraphics[scale=0.4]{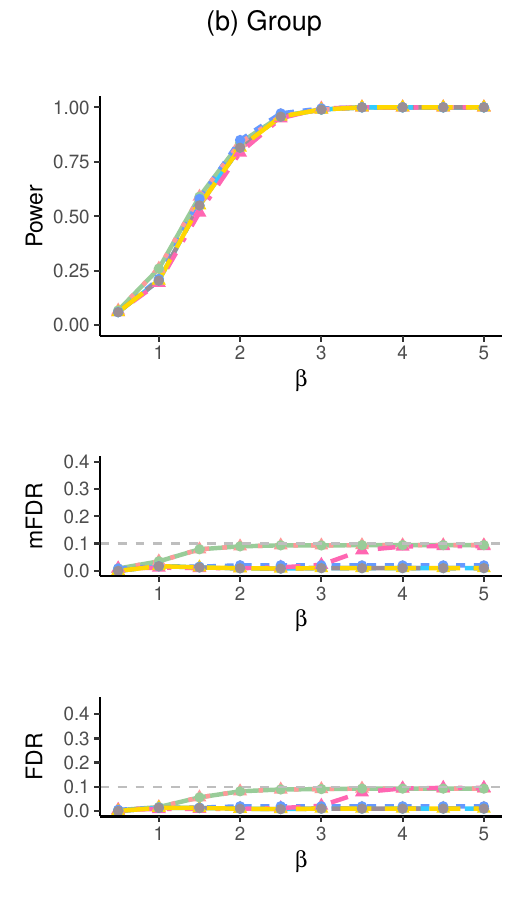}
     \includegraphics[scale=0.4]{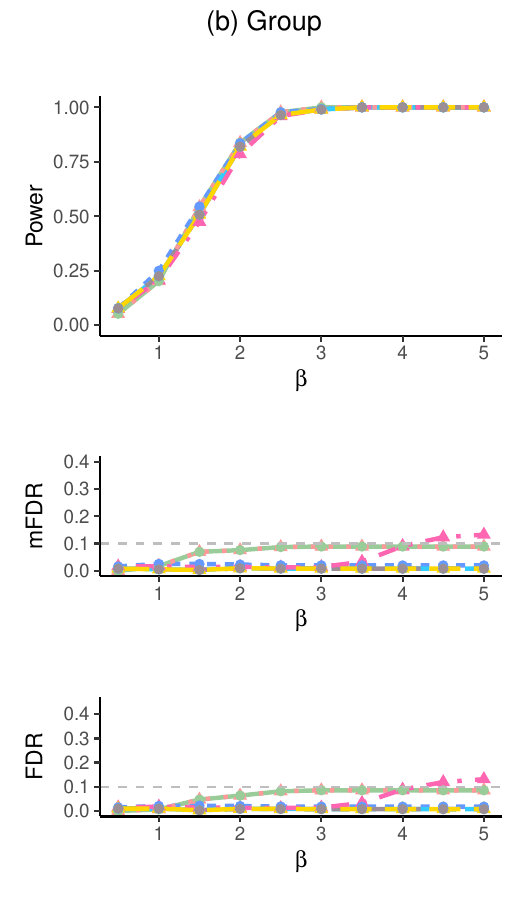}\\
   \caption{Power, FDR and mFDR for individual and group layers for different group structures with s = 20, k = 100, n = 10, G = 20, fixed signal pattern and constant signal strength: (a) Balanced Block Structure; (b) Balanced Interleaved Structure; (c) Unbalanced Structure}
   \label{fig:figure1}
\end{figure}

\begin{figure}[H]
    \centering
    \includegraphics[scale=1]{result/legend_file.pdf}\\
     \includegraphics[scale=0.4]{result/Figure_ind_100_2_1_1.pdf}
     \includegraphics[scale=0.4]{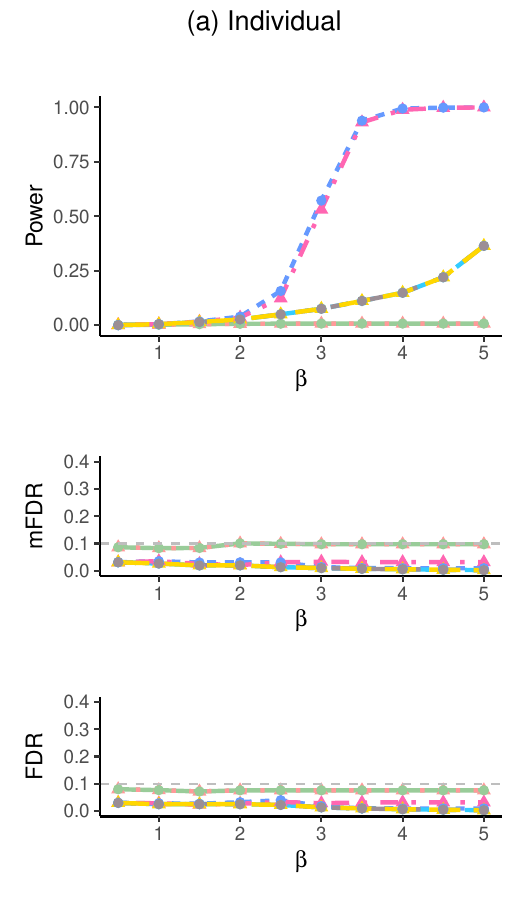}
     \includegraphics[scale=0.4]{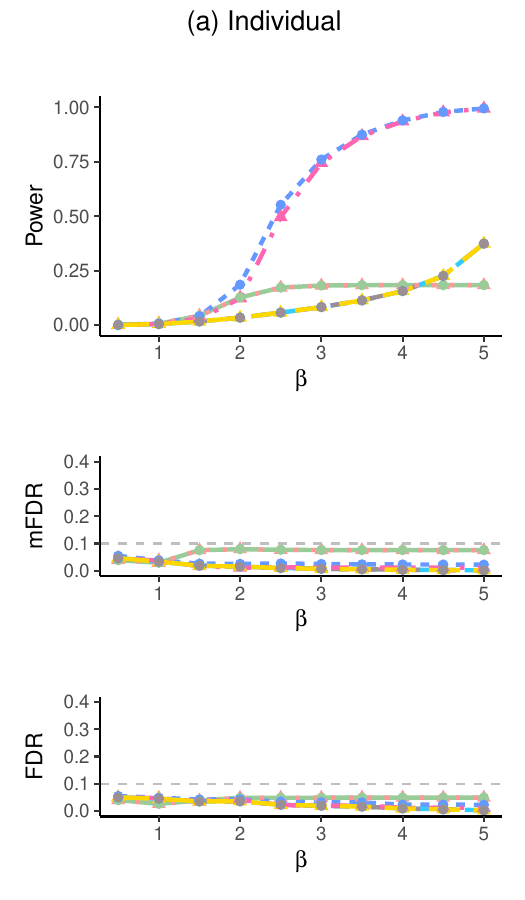}\\
     \includegraphics[scale=0.4]{result/Figure_group_100_2_1_1.pdf}
     \includegraphics[scale=0.4]{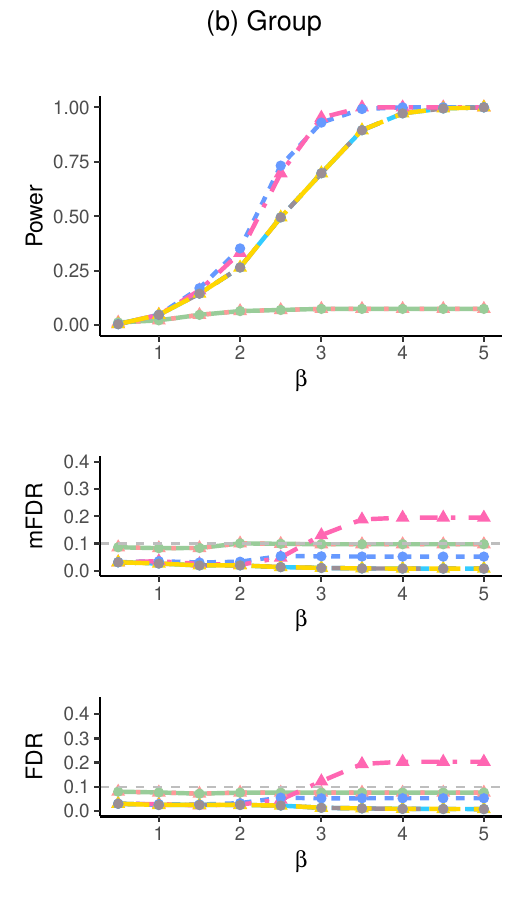}
     \includegraphics[scale=0.4]{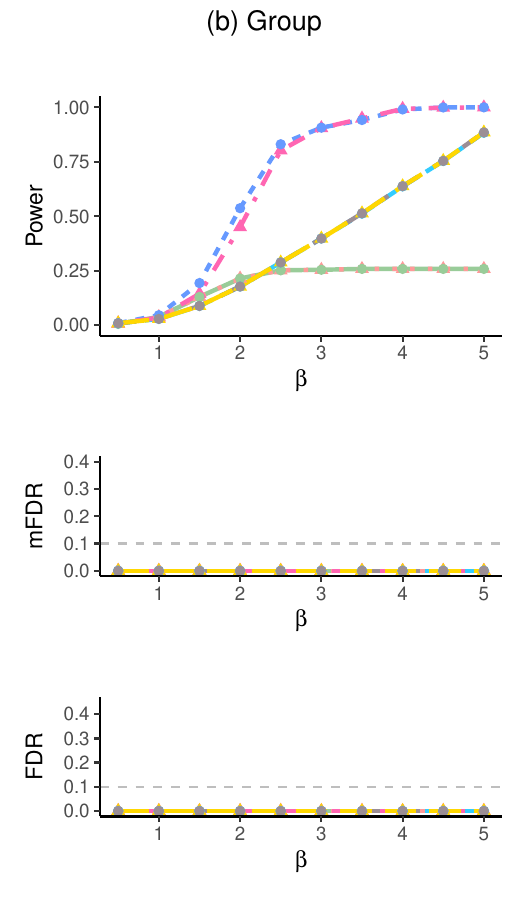}\\
  \caption{Power, FDR and mFDR for individual and group layers for different signal structure with s = 20, k = 100, n = 10 and G = 20, balanced interleaved group structure and constant signal strength: (a) Fixed pattern; (b) Random pattern; (c) Markov pattern}
   \label{fig:figure2}
\end{figure}

\begin{figure}[H]
    \centering
    \includegraphics[scale=1]{result/legend_file.pdf}\\
     \includegraphics[scale=0.4]{result/Figure_ind_100_1_1_1.pdf}
     \includegraphics[scale=0.4]{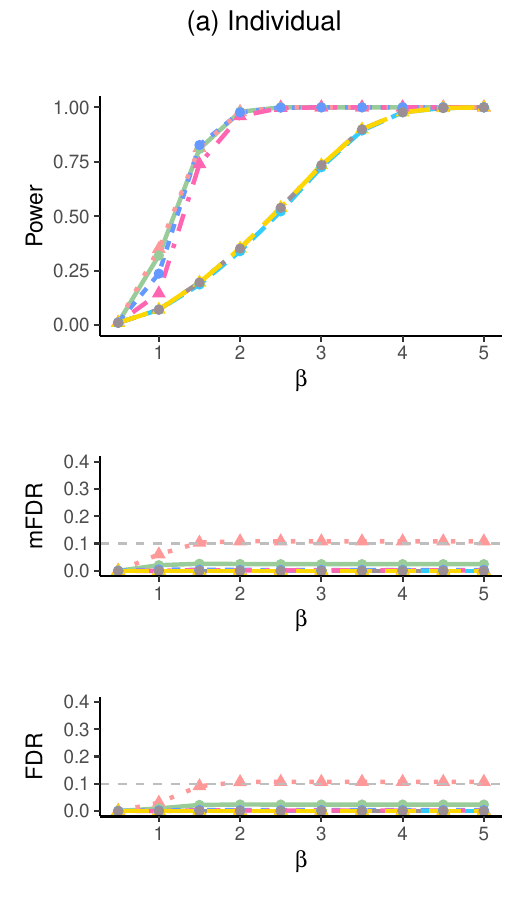}
     \includegraphics[scale=0.4]{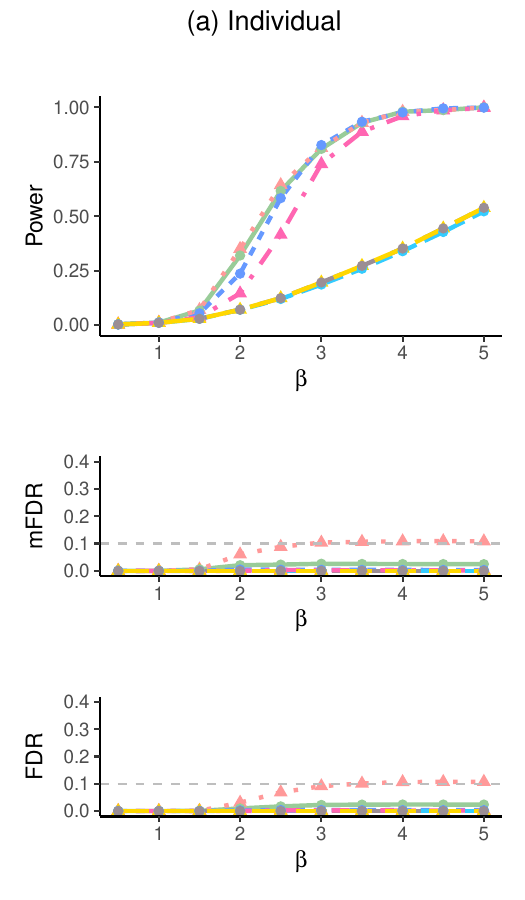}\\
     \includegraphics[scale=0.4]{result/Figure_group_100_1_1_1.pdf}
     \includegraphics[scale=0.4]{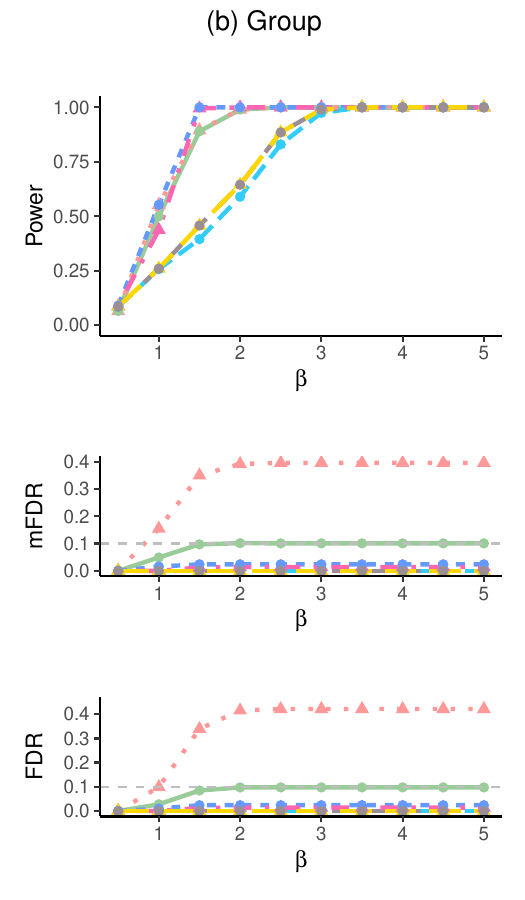}
     \includegraphics[scale=0.4]{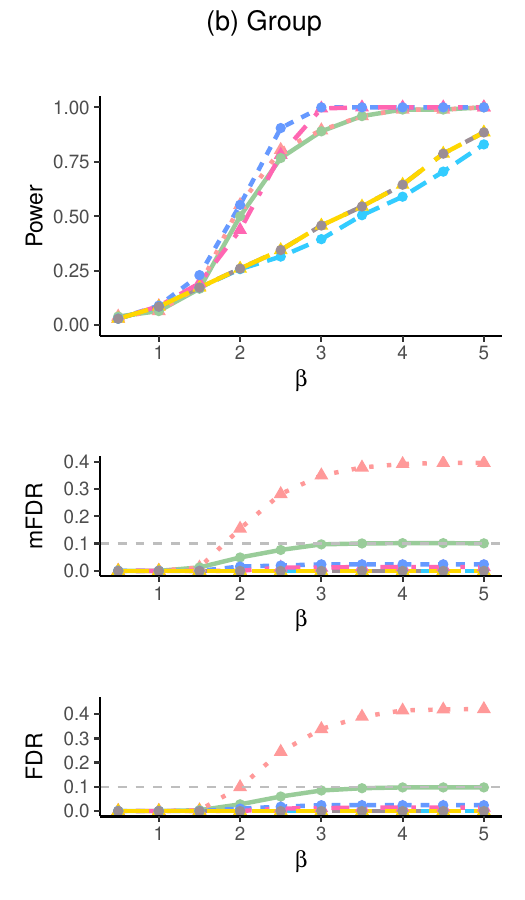}\\
   \caption{Power, FDR and mFDR for individual and group layers for different signal strength with s = 20, k = 100, n = 10, G = 20, balanced block group structure and fixed signal pattern: (a) Constant strength; (b) increasing strength; (c) Decreasing strength}
   \label{fig:figure3}
\end{figure}

\begin{figure}[H]
    \centering
     \includegraphics[scale=1]{result/legend_file.pdf}\\
     \includegraphics[scale=0.4]{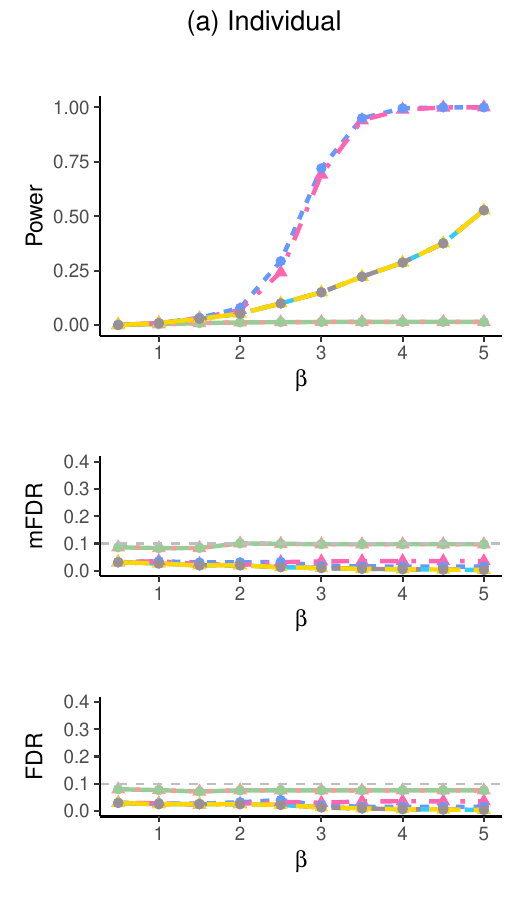}
     \includegraphics[scale=0.4]{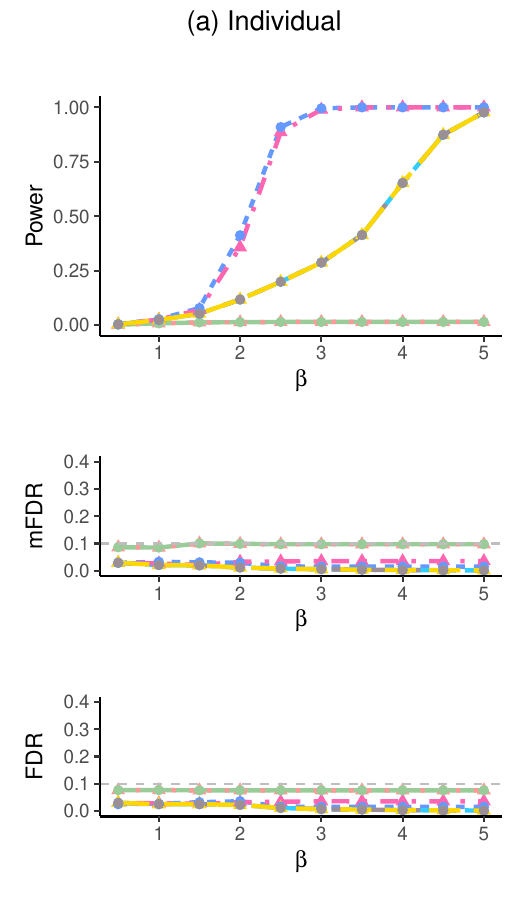}
     \includegraphics[scale=0.4]{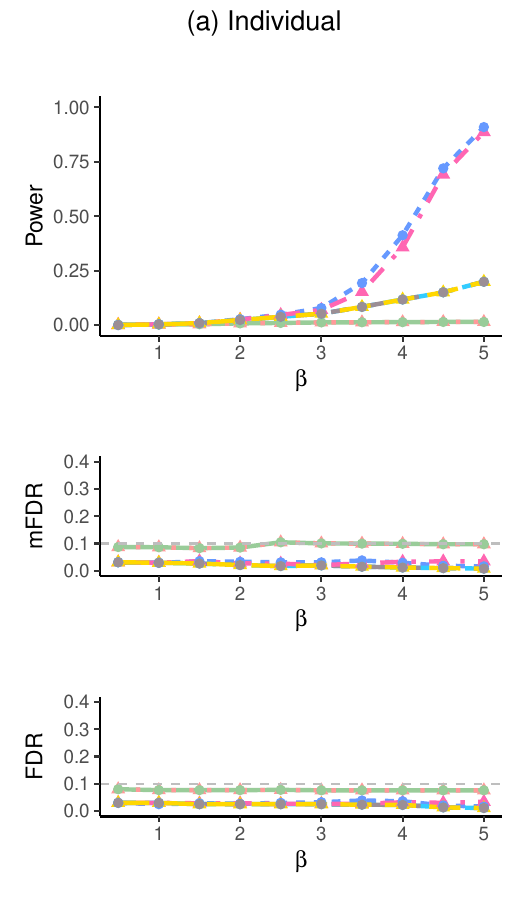}\\
     \includegraphics[scale=0.4]{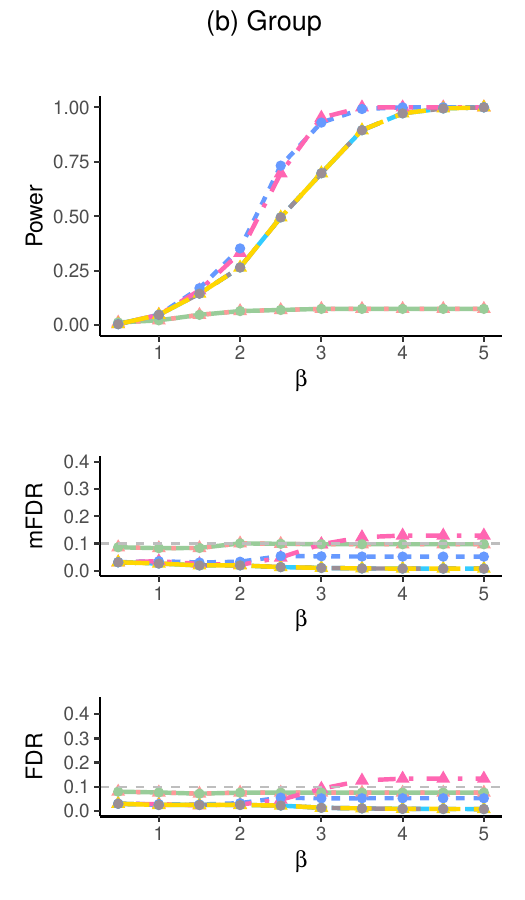}
     \includegraphics[scale=0.4]{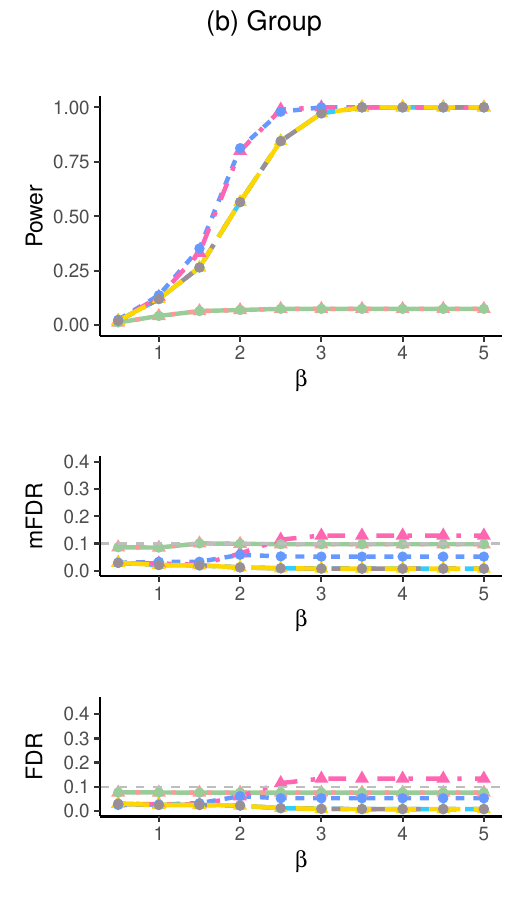}
     \includegraphics[scale=0.4]{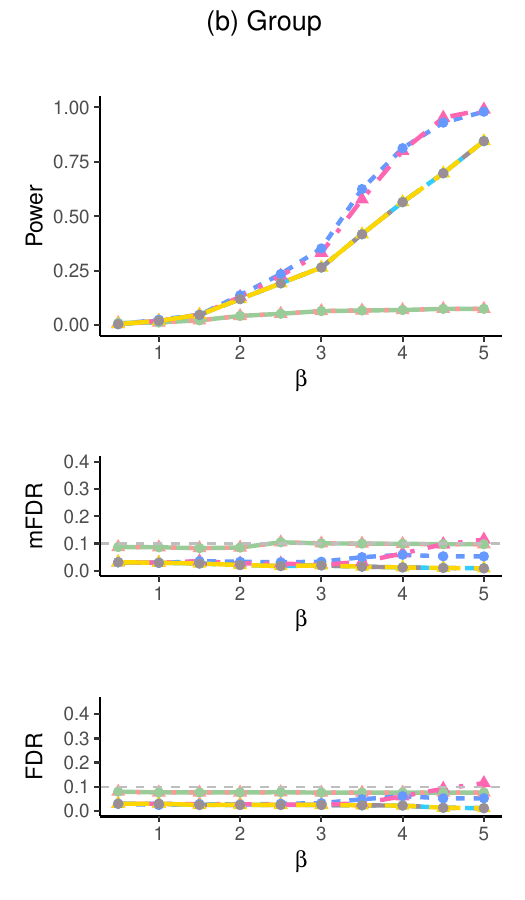}\\
   \caption{Power, FDR and mFDR for individual and group layers for different signal strength with s = 20, k = 50, n = 10, G = 20, balanced interleaved group structure and random signal pattern: (a) Constant strength; (b) increasing strength; (c) Decreasing strength}
   \label{fig:figure4}
\end{figure}

{\color{black} Our baseline configuration used the following parameters: 20\% true groups ($s = 20$), 50\% true signals within true groups ($k = 100$), 20 total groups ($G = 20$), 10 features per group ($n = 10$), balanced block structure fixed signal pattern, and constant signal strength. We varied the signal strength parameter $\beta$ to examine how the methods performed across different effect sizes.

Figure \ref{fig:figure1} shows the power, FDR, and mFDR for both individual and group layers under different group structures. For the \textbf{Balanced Block Structure}, most methods successfully maintained FDR and mFDR control at both individual and group levels except GAI. For GAI, both group-level FDR and mFDR exceed 0.4 when $\beta$ exceeds 2.0. The ml-LORD and LORD exhibit the highest power at both levels exhibit the highest power at both levels. As $\beta$ increases beyond 4.5, the power of all methods approaches 1.0. For the \textbf{Balanced Interleaved Structure}, we observe that LORD methods (both LORD and ml-LORD) significantly outperform all other methods in terms of power at both individual and group levels. Importantly, all multi-layer methods maintain FDR and mFDR control at both levels. The \textbf{Unbalanced Structure} demonstrates results largely consistent with \textbf{Balanced Interleaved Structure}. However, a slight exceedance of the 0.1 threshold is observed for the LORD procedure when $\beta$ reaches 0.5.

Figure \ref{fig:figure2} illustrates the performance metrics across different true signal patterns. All the methods control both individual and group FDR and mFDR in \textbf{Fixed pattern} and \textbf{Markov pattern}. However, LORD fails to control both group-level mFDR and FDR for \textbf{Random pattern} when $\beta$ exceeds 3. In terms of power, both LORD and ml-LORD have obviously better performance than other methods in the three patterns. 

Figure \ref{fig:figure3} shows the impact of different signal strength patterns on power and FDR control. The GAI method fails to control both group-level mFDR and FDR across the three patterns. In contrast, all multi-layer methods successfully control both mFDR and FDR. Regarding power, all versions of the LOND procedure (LOND, ml-LOND, and ml-LOND\_m) exhibit lower power compared to the $\alpha$-investing (GAI, ml-GAI) and LORD (LORD, ml-LORD) methods. Among the three patterns, the ml-LORD method demonstrates the best overall performance, as it consistently achieves the highest power while maintaining FDR control. Furthermore, Figure~\ref{fig:figure4} shows the results with a reduced number of true signals ($k = 50$). Most methods continue to control both FDR and mFDR. However, a slight exceedance of the 0.1 threshold is observed for the LORD method when the effect size $\beta$ exceeds 3.0.

Our extensive simulation studies highlight the advantages of multi-layer FDR control methods over their original ones. Under the \textbf{Balanced Block Structure}, the GAI procedure fails to maintain group-level FDR control. In more complex scenarios, such as the \textbf{Random Pattern} setting, the LORD method also fails to adequately control group-level FDR. In contrast, all multi-layer methods consistently maintain valid control of both FDR and mFDR across all settings. In terms of power, the modified multi-layer LORD method slightly outperforms the original LORD approach while maintaining comparable FDR control.

All the simulation results are consistent with theoretical results, showing that the multi-layer $\alpha$-investing method simultaneously controls mFDR at multiple layers {\color{black}(Theorem 1)}, and the multi-layer LOND and LORD procedures achieve simultaneous control of both FDR and mFDR across layers {\color{black}(Theorem 2 and 3)}. These findings demonstrate that multi-layer methods not only ensure more reliable FDR/mFDR control, particularly at the group level, but also deliver better statistical power compared to their original methods.
}

\section{Discussion}\label{discussion}
In summary, we propose three methods that simultaneously control mFDR or FDR for multiple layers of groups (partitions) of hypotheses in an online manner. The mFDR or FDR controls were theoretically guaranteed and demonstrated with simulations. These methods fill the important gap in drug discoveries when we need to control the FDR in drug screening for multiple layers of features (i.e. the siRNAs, nanocapsules, target cell lines etc.). {\color{black}For example, when screening siRNA-nanocapsule combinations, our methods allow researchers to make statements about both the efficacy of specific combinations and the general performance of specific siRNAs or nanocapsules across all combinations tested, which can accelerate the identification of promising candidates and reduce false leads. Besides, the proposed algorithms of the methods are also computationally efficient. The computational overhead of our multi-layer methods is minimal compared to original methods. This makes our methods highly scalable even for large-scale screening applications involving thousands of hypotheses across multiple layers.}

A challenge we face is that the powers of these methods reduces as the number of layers increase. We can potentially increase the power by adding more assumptions and estimating the pattern of the true signals and correlation between p-values more accurately. The modified multi-layer LOND method ($\text{ml-LOND}_m$) shows an effective example. It has better power performance than the multi-layer LOND in simulation; developing such methods with theoretical guarantees will be our future work.

{\color{black}Another limitation lies in the assumption of independence or weak dependence among p-values across different layers. In real-world settings, particularly in high-throughput biological studies or multi-modal drug screenings, the characteristics are often correlated. 
{\color{black}
Our multi-layer online frameworks do not assume the independence
across the multiple characteristics, because the key assumptions are based on the conditional distribution conditioning on the history information. The construction of the statistics for such scenarios is a practical challenge. Some extensions of conditional permutation test \cite{berrett2020conditional} can be explored as a future direction.
}
}

\bibliography{example_paper}
\bibliographystyle{icml2022}


\pagebreak
\section*{Web Appendix A: Technical Proof}  \label{app:proof}

\begin{proof}[Proof of Lemma 1]
Denote $\mathbb{E}_{\theta}^{j-1} = \mathbb{E}_{\theta}(\cdot|\mathcal{F}_{j-1})$. The key here is that $\phi_j^m$, $\psi_j^m$ and $\alpha_j^m$ are measurable with respect to $\mathcal{F}_{j-1}$. For a layer $m$ that we need to test at time $j$, we can deduce that $\delta_{g_j^m}^m(j - 1) = 0$. If $\theta_j = 0$ and $\theta_{g_j^m}^m(j - 1) = 0$, then $\theta_{g_j^m}^m(j) = 0$ and $V^m(j) - V^m(j - 1) = \delta_{g_j^m}^m(j)$, we have
\begin{align*}
\mathbb{E}_{\theta}^{j-1}(A^m(j) - A^m(j - 1)) &= -(1 - \alpha + \psi_j^m)\mathbb{E}_{\theta}^{j-1}(\delta_{g_j^m}^m(j)) + \phi_j^m \\
&\geq -(1 - \alpha + \psi_j^m)\alpha_j^m + \phi_j^m \\
&\geq -(1 - \alpha + \phi_j^m/\alpha_j^m + \alpha - 1)\alpha_j^m + \phi_j^m = 0
\end{align*}

If $\theta_j = 0$ and $\theta_{g_j^m}^m(j -1) = 1$, then $\theta_{g_j^m}^m(j) = 1$ and $V^m(j)-V^m(j -1) = 0$, we have
\begin{align*}
\mathbb{E}_{\theta}^{j-1}(A^m(j) - A^m(j - 1)) &= -(-\alpha + \psi_j^m)\mathbb{E}_{\theta}^{j-1}(\delta_{g_j^m}^m(j)) + \phi_j^m \\
&\geq -(-\alpha + \psi_j^m)\alpha_j^m + \phi_j^m \\
&\geq -(-\alpha + \phi_j^m/\alpha_j^m + \alpha - 1)\alpha_j^m + \phi_j^m = \alpha_j^m \geq 0
\end{align*}

If $\theta_j = 1$, then $\theta_{g_j^m}^m(j) = 1$ and $V^m(j) - V^m(j - 1) = 0$, we have
\begin{align*}
\mathbb{E}_{\theta}^{j-1}(A^m(j) - A^m(j - 1)) &= -(-\alpha + \psi_j^m)\mathbb{E}_{\theta}^{j-1}(\delta_{g_j^m}^m(j)) + \phi_j^m \\
&\geq -(-\alpha + \psi_j^m)\rho_j^m + \phi_j^m \\
&\geq -(-\alpha + \phi_j^m/\rho_j^m + \alpha)\rho_j^m + \phi_j^m = 0
\end{align*}
This finishes the proof that $A^m(j)$ is a submartingale.
\end{proof}

\begin{proof}[Proof of Theorem 1]
Since $R^m(0) = 0$, $V^m(0) = 0$, $W^m(0) = \alpha\eta$, so we have $A^m(0) = 0$. By Lemma 1, we have $\mathbb{E}_{\theta}(A^m(t)) = \alpha\mathbb{E}_{\theta}R^m(t) - \mathbb{E}_{\theta}V^m(t) + \alpha\eta - \mathbb{E}_{\theta}W^m(t) \geq 0$. Since $W^m(t) \geq 0$, $\mathbb{E}_{\theta}(A^m(t)) = \alpha\mathbb{E}_{\theta}R^m(t) - \mathbb{E}_{\theta}V^m(t) + \alpha\eta \geq 0$ which lead to $mFDR_{\eta}^m = \frac{\mathbb{E}_{\theta}V^m(t)}{\mathbb{E}_{\theta}R^m(t)+\eta} \leq \alpha$ as long as $t$ is a stopping time.
\end{proof}

\begin{proof}[Proof of Theorem 2]
We first show that $FDR^m(n) \leq \alpha$. Write
\begin{align*}
FDR^m(n) &= \mathbb{E}_{\theta}\left(\frac{V^m(n)}{R^m(n) \vee 1}\right) \\
&\leq \mathbb{E}_{\theta}\left(\sum_{j=1}^n \frac{\delta_{g_j^m}^m(j)}{R^m(n) \vee 1}\right) \\
&= \mathbb{E}_{\theta}\left(\sum_{j=1}^n \frac{\delta_{g_j^m}^m(j)}{\alpha_j^m} \cdot \frac{\alpha_j^m}{R^m(n) \vee 1}\right)
\end{align*}

Since $R^m(n) \vee 1 \geq R^m(j) \vee 1$, we have
\[
\frac{\alpha_j^m}{R^m(n) \vee 1} \leq \frac{\alpha_j^m}{R^m(j - 1) + 1} = \beta_j^m
\]
and by assumption, we know that $\forall \theta \in \Theta$, $\forall m = 1, \cdots, M$, we have 
\[
\mathbb{E}_{\theta}(\delta_{g_j^m}^m(j) - \alpha_j^m|R^m(j - 1), \delta_{g_i^m}^m(i - 1) = 0) \leq 0.
\]

So we have
\[
\mathbb{E}_{\theta}\left(\frac{\delta_{g_j^m}^m(j)}{\alpha_j^m}\right) \leq \mathbb{E}_{\theta}\left(\frac{\delta_{g_j^m}^m(j)}{\alpha_j^m}|R^m(j - 1), \delta_{g_i^m}^m(i - 1) = 0\right) \leq \mathbb{E}(1) = 1
\]
since $\alpha_j^m$ is measurable with respect to $R^m(j - 1)$. Summing over $n$, we obtain $FDR^m(n) \leq \sum_{j=1}^{\infty} \beta_j^m = \alpha$.

We next prove that $mFDR^m(n) \leq \alpha$. Since
\[
\mathbb{E}_{\theta}(\delta_{g_i^m}^m - \alpha_j^m) = \mathbb{E}[\mathbb{E}_{\theta}(\delta_{g_i^m}^m - \alpha_j^m|R^m(j - 1), \delta_{g_i^m}^m(i - 1) = 0)] \leq 0.
\]

Summing over $n$, we have
\begin{align*}
\mathbb{E}_{\theta}(V^m(n)) &\leq \sum_{i=1}^n \mathbb{E}_{\theta}\alpha_i^m \\
&= \sum_{i=1}^n \mathbb{E}_{\theta}(\beta_i^m R^m(i - 1) + 1) \\
&= \sum_{i=1}^n \mathbb{E}_{\theta}\left(\beta_i^m \sum_{j=1}^{i-1} \delta_j^m\right) + \sum_{i=1}^n \beta_i^m \\
&= \sum_{j=1}^{n-1} \left(\sum_{i=j+1}^n \beta_i^m\right)\mathbb{E}_{\theta}(\delta_i^m) + \sum_{i=1}^n \beta_i^m \\
&\leq \left(\sum_{i=1}^n \beta_i^m\right)(\mathbb{E}_{\theta}(R^m(n)) + 1)
\end{align*}

So $mFDR^m(n) \leq \sum_{i=1}^n \beta_i^m \leq \alpha$.
\end{proof}

\begin{proof}[Proof of Theorem 3]
Because the sum of significance level $\alpha_i^m$ between two consecutive discoveries is bounded by $\alpha$, so
\[
\mathbb{E}_{\theta}(V^m(n)) \leq \sum_{i=1}^n \mathbb{E}_{\theta}\alpha_i^m \leq \alpha\mathbb{E}_{\theta}(R^m(n) + 1)
\]

This leads to the result that $mFDR_1^m(n) \leq \alpha$.

Next, we prove the result for FDR. Denote $X_i^m = 1$ if $d_i^m < \infty$ and the $i$-th discovery is false, $X_i^m = 0$ if $d_i^m < \infty$ and the $i$-th discovery is true and $X_i^m = 2$ otherwise. First we have
\[
\mathbb{E}_{\theta}(X_i^m I(d_i^m < \infty)|\mathcal{F}_{d_{i-1}^m}) \leq \sum_{l=d_{i-1}^m+1}^{\infty} P_{\theta_l=0}(\delta_{g_l^m}^m(l) = 1, d_i^m = l|\mathcal{F}_{d_{i-1}^m})
\]

Given $d_{i-1}^m$, we know $\zeta_i^m = \kappa_i^m - \kappa_{d_{i-1}^m}^m + 1$. This implies that for any $l = d_{i-1}^m + 1, \cdots, \infty$, $\zeta_l^m = \zeta_{l-1}^m$ is equivalent to $\kappa_l^m = \kappa_{l-1}^m$, which means $l$ will not be tested in the layer $m$ and it is impossible for $d_i^m = l$. So we have
\begin{align*}
\sum_{l=d_{i-1}^m+1}^{\infty} P_{\theta_l=0}(\delta_{g_l^m}^m(l) = 1, d_i^m = l|\mathcal{F}^{m}_{d_{i-1}^m}) \\
= \sum_{l=d_{i-1}^m+1}^{\infty} P_{\theta_l=0}(\delta_{g_l^m}^m(l) = 1, d_i^m = l|\mathcal{F}^{m}_{d_{i-1}^m})I(\zeta_i^m \neq \zeta_{i-1}^m) \\
\leq \sum_{l=d_{i-1}^m+1}^{\infty} \alpha_i^m I(\zeta_i^m \neq \zeta_{i-1}^m) \leq \sum_{r=1}^{\infty} \beta_r^m = \alpha
\end{align*}
The last part holds since $\alpha_i^m = \beta_{\zeta_i^m}^m$.

\begin{align*}
\mathbb{E}_{\theta}(FDP(d_k^m)I(d_i^m < \infty)) &= \mathbb{E}\left[\frac{V^m(d_k^m)}{k}I(d_i^m < \infty)\right] \\
&= \frac{1}{k}\sum_{i=1}^k \mathbb{E}(X_i^m I(d_i^m < \infty)) \\
&\leq \frac{1}{k}\sum_{i=1}^k \mathbb{E}[\mathbb{E}(X_i^m I(d_i^m < \infty))|\mathcal{F}_{d_{i-1}^m}] \leq \alpha
\end{align*}
\end{proof}
\end{document}